\documentstyle[11pt,twoside,colloqOHP2013,epsfig,
 natbib]{article}
\begin{document}

\def\josa{Jour. Opt. Soc. America}%
\def\aj{AJ}%
\def\actaa{Acta Astron.}%
\def\araa{ARA\&A}%
\def\apj{ApJ}%
\def\apjl{ApJ}%
\def\apjs{ApJS}%
\def\ao{Appl.~Opt.}%
\def\apss{Ap\&SS}%
\def\aap{A\&A}%
\def\aapr{A\&A~Rev.}%
\def\aaps{A\&AS}%
\def\azh{AZh}%
\def\baas{BAAS}%
\def\bac{Bull. astr. Inst. Czechosl.}%
\def\caa{Chinese Astron. Astrophys.}%
\def\cjaa{Chinese J. Astron. Astrophys.}%
\def\icarus{Icarus}%
\def\jcap{J. Cosmology Astropart. Phys.}%
\def\jrasc{JRASC}%
\def\mnras{MNRAS}%
\def\memras{MmRAS}%
\def\na{New A}%
\def\nar{New A Rev.}%
\def\pasa{PASA}%
\def\pra{Phys.~Rev.~A}%
\def\prb{Phys.~Rev.~B}%
\def\prc{Phys.~Rev.~C}%
\def\prd{Phys.~Rev.~D}%
\def\pre{Phys.~Rev.~E}%
\def\prl{Phys.~Rev.~Lett.}%
\def\pasp{PASP}%
\def\pasj{PASJ}%
\def\qjras{QJRAS}%
\def\rmxaa{Rev. Mexicana Astron. Astrofis.}%
\def\skytel{S\&T}%
\def\solphys{Sol.~Phys.}%
\def\sovast{Soviet~Ast.}%
\def\ssr{Space~Sci.~Rev.}%
\def\zap{ZAp}%
\def\nat{Nature}%
\def\iaucirc{IAU~Circ.}%
\def\aplett{Astrophys.~Lett.}%
\def\apspr{Astrophys.~Space~Phys.~Res.}%
\def\bain{Bull.~Astron.~Inst.~Netherlands}%
\def\fcp{Fund.~Cosmic~Phys.}%
\def\gca{Geochim.~Cosmochim.~Acta}%
\def\grl{Geophys.~Res.~Lett.}%
\def\jcp{J.~Chem.~Phys.}%
\def\jgr{J.~Geophys.~Res.}%
\def\jqsrt{J.~Quant.~Spec.~Radiat.~Transf.}%
\def\memsai{Mem.~Soc.~Astron.~Italiana}%
\def\nphysa{Nucl.~Phys.~A}%
\def\physrep{Phys.~Rep.}%
\def\physscr{Phys.~Scr}%
\def\planss{Planet.~Space~Sci.}%
\def\procspie{Proc.~SPIE}%

%
%
\title{DISCO: a Spatio-Spectral Recombiner for Pupil Remapping Interferometry}
\author{F. Millour$^{1}$, R. Petrov$^1$, S. Lagarde$^1$, P. Berio$^1$, Y. Bresson$^1$, L. Abe$^1$}
\affil{$^1$Laboratoire Lagrange, UMR7293, Universit\'e de Nice Sophia-Antipolis, CNRS, Observatoire de la C\^ote dÕAzur, Bd. de l'Observatoire, 06304 Nice, France}

\begin{abstract}
Pupil-remapping is a new high-dynamic range imaging technique that has recently demonstrated feasibility on sky. The current prototypes present however deceiving limiting magnitude, restricting the current use to the brightest stars in the sky.
We propose  to combine pupil-remapping with spatio-spectral encoding, a technique first applied to the VEGA/CHARA interferometer. The result is an instrument proposal, called "Dividing Interferometer for Stars Characterizations and Observations" (DISCO). The idea is to take profit of wavelength multiplexing when using a spectrograph in order to pack as much as possible the available information, yet providing a potential boost of 1.5 magnitude if used in existing prototypes.
We detail in this paper the potential of such a concept.
\end{abstract}
\section{Introduction}

The need for a better dynamic range in direct imaging techniques is today identified as a top priority for the detection and characterization of extrasolar planets. As an illustration, several high-dynamic range imaging instruments are currently being developed \citep[notably: SPHERE, HICIAO or GPI / ][]{Beuzit2006, 2006dies.conf..323T, 2006SPIE.6272E..18M}. These instruments make use of so-called "Extreme-Adaptive Optics" (XAO) in order to make coherent (i.e. interfering) the highest number of photons in the resulting image.

An other way existed before the advent of adaptive optics to get coherent photons: speckle imaging \citep{1970A&A.....6...85L, 1978otf..conf..479L} makes use of short-integration times to freeze the Earth's atmosphere disturbance and take over its resolution-washing effect. However, the speckle technique and all of its derivatives (speckle masking, segment tilting, lucky imaging, etc.) are bound to waste photons in a way or in another. This is why pupil remapping was proposed  by \cite{Perrin2006a}, to take profit of both fully coherent photons and full-pupil flux collection.

Since the original idea was proposed, pupil remapping has evolved from a pure concept up to a readily demonstrated instrument on-sky \citep{Huby2012, Huby2013}. The built prototypes have shown the great potential of this technique and also some limitations.

\begin{figure}[htbp]
\begin{center}
\epsfig{file=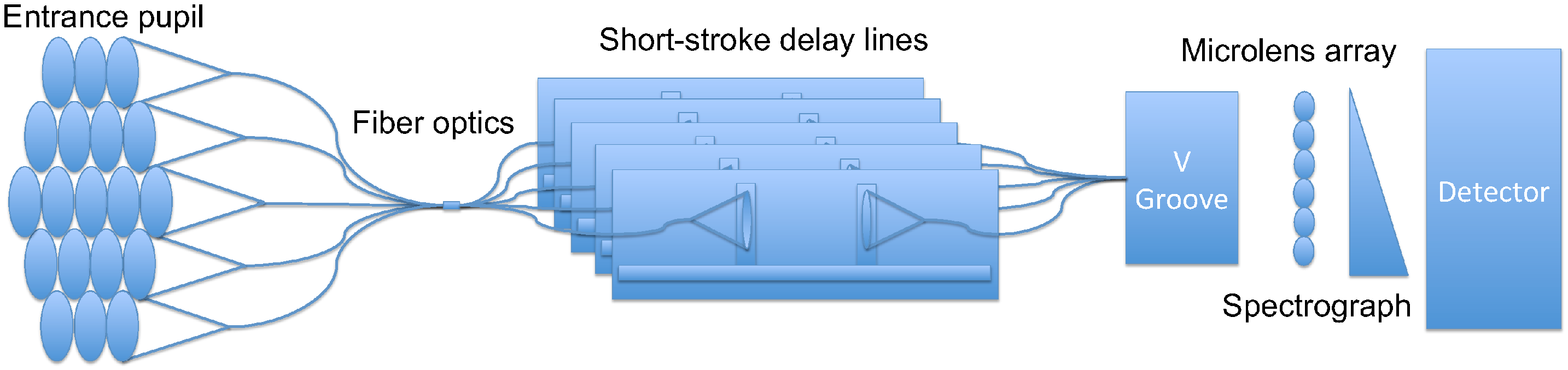,height=11cm,angle=-90}
\caption{A possible setup for DISCO. Please note the similarities with the sketch in \cite{Perrin2006a}. The differing parts are the short-stroke delay lines and the arrangement of fibers in the V groove, described in the current paper.}
\label{fig:sketchInstrument}
\end{center}
\end{figure}

Here, we propose an improvement over the pupil remapping concept as presented in \cite{Huby2012}, in order to collect more photons per pixel for a given setup. While this might sound useless for some applications where pixels are "cheap" (like in visible applications), IR wavelength detectors still have limitations on their detector readout noise, making each pixel valuable. A sketch of such an instrument is presented in Fig.~\ref{fig:sketchInstrument}, which is similar to the proposal of \cite{Perrin2006a}. It differs mainly in the addition of short-stroke delay lines to control the optical path difference (hereafter OPD) and in the output pupil configuration, which is described in the next section.

We will therefore briefly describe how do we plan to save on pixels, and present optimized OPD configurations to use in such an instrument.

\section{A recall of the technique and proposal of a new scheme}

In an all-in-one multi axial interferometer, several beams are combined altogether, coding the fringes by their frequencies. One baseline corresponds then to one spatial frequency (a "fringe peak") in the Fourier Transform (FT) of the fringe pattern.

It has long been theorized that only a fully non-redundant configuration would allow one to extract the interferometric signal. Hence, several instrument were built on such a beam configuration: the AMBER \citep{2007A&A...464....1P}, or MIRC \citep{2004SPIE.5491.1370M} combiner are a few examples.

However, it was proposed in the first times of optical long-baseline interferometry \citep{Vakili1989}, any more recently demonstrated on a wider scale with the VEGA instrument \citep{Mourard2011}, that a fully redundant configuration could also be used given that the fringes could be spectrally dispersed with a sufficient spectral resolution. In such a case, the fringe peaks of several baselines are at the same spatial frequency, noted $V_{pi}$  in order to take the same notation as in \cite{Mourard2011}, making them totally cluttered in usual analysis algorithms. However, they can be disentangled by inputting a different fixed OPD, which in turn allows one to change the peaks positions in the wavelength frequency domain, noted $U_{pi}$ in \cite{Mourard2011}. A different approach for data analysis has to be used, with the use of 2D FTs instead of 1D FTs, which is extensively described in \cite{Mourard2011}. An additional way of uncluttering the fringe peaks is to input an OPD modulation on groups of sub-apertures and to make use of 3-dimensional Fourier Transforms (the third dimension being along time), as was proposed by \citet{Vakili1989}.

\begin{figure}[htbp]
\begin{center}
\epsfig{file=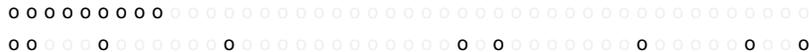,width=11cm}
\caption{On top is the output pupil of a fully redundant configuration. On the bottom is the non-redundant configuration used in the prototype FIRST instrument.}
\label{fig:redundantvsnonredundant}
\end{center}
\end{figure}

We reproduce in Fig.~\ref{fig:redundantvsnonredundant} the 9 sub-apertures non-redundant output pupil used in the FIRST instrument \citep{Huby2012}, and side to side, the output pupil of a fully redundant configuration.

Such a configuration saves a great deal of pixels compared to non-redundant configuration, with a given number of sub-apertures and spectral resolution. In the case of 9 sub-apertures, one can save a factor 5, i.e. a direct $\approx1.5$ magnitude gain given the same spectral resolution.

We evaluate in the next section the requirements and identified limits of such an instrument.

\section{Optimizing a spatio-spectral interferometer}

\subsection{Spectral resolution}

When dealing with fully redundant output pupil for an interferometer, one needs to set a minimum OPD distance between the peaks in order to avoid peaks overlap. On the other hand, the applied OPDs must not exceed a fraction of the coherence length of the fringe pattern, otherwise loss of contrast and under sampling effects would occur.

These conditions provide guidelines that will set the range of spectral resolution and the values of OPD offsets to use in such an instrument.

\subsubsection{Minimal condition}

We recall  the minimal condition on the number of spectral channels to use, given in \citet[][equation 13]{Mourard2011}:

\begin{equation}
N_{\rm ch} \geq 2 N_{\rm tel}
\end{equation}

This puts a condition on the minimum spectral resolution to use:

\begin{equation}
R \geq 2 N_{\rm tel} \frac{\lambda_0}{\Delta \lambda}
\label{eq:condition}
\end{equation}

with $\lambda_0$ the central wavelength of the observations, and $\Delta \lambda$ the observation bandwidth.
For a 9 sub-apertures instrument, working in the K band ($\lambda_0=2.2\mu$m, $\Delta\lambda=0.4\mu$m), this imposes a minimum spectral resolution of $\approx100$.

However, this is a conservative limit, as there are $N_{\rm tel}-1$ fringe peaks for a given frequency, which hence can be compared with $N_{\rm tel}-2$ minimal distances. We call CDR the ratio between the largest necessary OPD to input for one given configuration and the shortest distance between two adjacent peaks.

This CDR can be optimized but is always by construction greater or equal to $N_{\rm tel}-2$ and depends on the number of sub-apertures used. 
In the example given above, for a 9-subapertures configuration, we get a CDR of 7 (see Table~\ref{tab:optimizedValues}). This translates into a minimum spectral resolution of  $\approx80$.

When optimizing the configuration (section \ref{sec:opdOffsets}), we can see that this CDR can be used as a criterion to minimize, in order to pack as much as possible the fringe peaks together.

\subsubsection{Atmosphere and/or adaptive optics jitter}

Another criterion to consider is the wobbling of the fringe peaks by natural-atmospheric or adaptive optics-induced OPD. The peaks separation in the $U$ direction must be greater than twice the atmospheric wobbling. If we consider the atmospheric OPD over Paranal which has a peak overrun ${\rm OPD}_{\rm max}$ of typically $25\mu$m \citep{2007A&A...464...29T}, this means that two adjacent fringe peaks must be separated typically by $50\mu$m.

This imposes conditions on the coherence length $L_c$, that must follow the condition:

\begin{equation}
L_c \geq 2 {\rm CDR} \times {\rm OPD}_{\rm max}
\label{eq:conditionOPD}
\end{equation}

or 

\begin{equation}
R \geq \frac{2 {\rm CDR} \times {\rm OPD}_{\rm max}}{\lambda_0}
\label{eq:conditionOPD}
\end{equation}

So, still for the 9-telescope configuration example given above, the minimum spectral resolution to use would be  $\approx160$. We see that in such a configuration, the fringes wandering by the atmosphere is by far the most stringent constrain on the spectral resolution. However, the use of adaptive optics prior to the input pupil (by reducing the OPD wandering from $25\mu$m to less than $1\mu$m), or the use of  OPD modulation proposed in \citet{Vakili1989} could strongly relax this constrain.

\subsubsection{Fringe peaks overlap}

As was highlighted by J. Monnier during the conference, an overlap of the fringe peaks could occur due to the spectrum shape of the target. Two ways of overcoming this effect were presented in \citet{Mourard2011} by using differential measurements combined with either setting a minimal width of the work channel, or by solving a set of equation describing the peaks overlap.

It is worth to mention that partial peaks overlap could also occur in non-redundant configurations, as happens in the AMBER instrument \citep{2004SPIE.5491.1222M, 2007A&A...464...29T}. The use of an image-based algorithm (the P2VM) solves this issue, and one could consider also using a 2D-image-based model-fitting algorithm, similar to the P2VM, to avoid the peaks contamination in our case.

\subsection{OPD offsets optimization}
\label{sec:opdOffsets}

In the literature, a few papers consider the problem of optimizing frequencies in an array. We can cite for example \cite{Moffet1968e, Vertatschitsch1986, Ribak1988a, Pearson1990} for aligned sub-apertures with or without some redundancy, and \cite{Golay1971} for 2D optimization. However, we found no trace of spatio-spectral optimization, except in the two papers \citet{Vakili1989, Mourard2011} where setups for specific configurations were provided.

In \cite{Mourard2011} are addressed the cases of 3 and 4 telescopes for the spatio-spectral instrument VEGA. In \cite{Vakili1989} is presented an example with 12 telescopes. Since we discuss  the possibility to combine tens of sub-apertures for a potential full-pupil instrument, we investigated the optimization of the spatio-spectral scheme for up to 64 sub-apertures, though we present here only a subset, up to 30 sub-apertures.

We considered for this optimization the minimization of the CDR, in order to separate the peaks at maximum. We define this new criterion instead of using moment of inertia or other criteria defined in \citet{Golay1971} because though we end up with 2D fringe peaks patterns, we aim at only optimizing one dimension (the OPD dimension).

We made use of a Monte-Carlo approach similar as in \cite{Ribak1988a}, using a  simulated annealing algorithm. Indeed, the number of fringe peaks for a given configuration scales as $N_{\rm tel}^2$, so the number of distances between fringe peaks to optimize scales as $N_{\rm tel}^4$. Therefore simple optimization methods like gradient descent would fail in finding an acceptable solution.

Table \ref{tab:optimizedValues} shows the results of our optimization for up to 9 sub-apertures with the corresponding CDR and minimum spectral resolution to use given an uncorrected atmosphere similar to Paranal. Interestingly, these 7 configurations  happen to have exactly ${\rm CDR} = N_{\rm tel}-2$, i.e. there exist no configurations more compact for these numbers of sub-apertures (though there exist other configurations with the same compactness, in which case we select the configuration with the least number of high-value OPD). The corresponding OPDs are given in $\mu$m for an instrument working in the K-band ($\lambda_0=2.2\mu$m, $\Delta\lambda=0.4\mu$m). We see that a even a moderate spectral resolution of $\approx160$ can be used to combine 9 telescopes.

\begin{table}[htbp]
\caption{7 most compact optimized OPDs for different interferometer configurations. We provide also the minimum required spectral resolution to avoid fringe peak overlap under a Paranal-like atmosphere in the K-band, and provide the OPDs for this spectral resolution (in $\mu$m)}
\label{tab:optimizedValues}
\small
\centering
\begin{tabular}{l*{35}{c}}
\tableline
&&&
\multicolumn{8}{c}{Beam \# OPD offset}\\
$N_{\rm tel}$ & CDR & $R_{\rm min}$  &
 1&2&3&4&5&6 &7&8 &9   \\
 \tableline
        3 & 1 & 33 & -36 & 36 &  -36\\
        4 & 2 & 45 & -50 & 50 & -50 &  -50\\
        5 & 3 & 68 & 75 & -75 & -25 & -75 &  75\\
        6 & 4 & 91 & -100 & 100 & 0 & 100 & -100 &  -100\\
        7 & 5 & 114 & 25 & -125 & 125 & 75 & 125 & -125 &  25\\
        8 & 6 & 136 & 150 & 50 & -150 & -150 & -50 & 150 & -150 &  150\\
        9 & 7 & 159 & -25 & 125 & 175 & -175 & 75 & -175 & 175 & 125 &  -25\\
\tableline
\end{tabular}
\end{table}

\begin{figure}[htbp]
\begin{center}
\epsfig{file=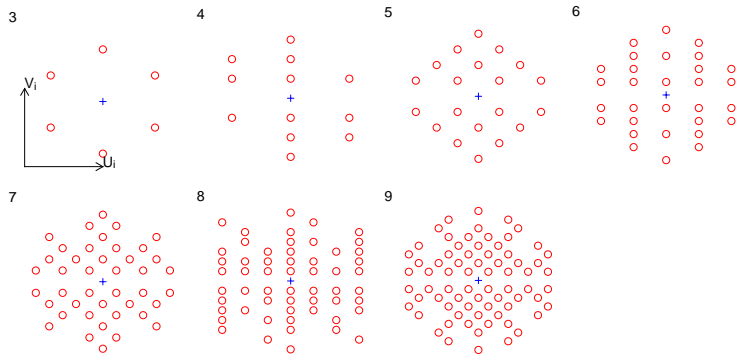,height=10cm,angle=-90}
\caption{The UV fringe peaks relative positions of the most compact optimized OPDs of Table~\ref{tab:optimizedValues}.}
\label{fig:UVpeaks}
\end{center}
\end{figure}

Figure~\ref{fig:UVpeaks} illustrates the appearance of the 2D Fourier transform by materializing the positions of the fringe peaks for 3 to 9 sub-apertures.

We also note that these given offset can be set as fixed OPDs, but can also be set as fringe drift speeds, if one considers a fully redundant interferometer with OPD modulation. In such case, instead of inputting fixed OPDs and analyzing the data as a function of wavelength, one can input OPD drifts with drifting speed proportional to the values in Table~\ref{tab:optimizedValues}, and analyze the data as a function of time. The great advantage of this alternate solution is to allow for a broadband instrument to be setup. A detailed analysis of such a concept is out of the scope of the current paper.

\subsection{avoiding zero OPD}
\label{sec:opdOffsets}

We see in Fig.~\ref{fig:UVpeaks} that up to 6 fringe peaks can be exactly at OPD 0 (for the 8 telescope configuration), which in some cases can be annoying due to the diffraction spike of the zero-frequency photometric peak. A way of overcoming such an issue is to input additional OPD offsets to the ones  provided here, which are proportional to the sub-aperture number. Such additional OPD offsets "skew" the peaks position sketch shown in Fig~\ref{fig:UVpeaks} and move all the central fringe peaks away from the zero OPD. Such an additional offset degrades slightly the CDR of the configuration. For example, for sub-aperture, one would need to add to the values of Table~\ref{tab:optimizedValues}, OPD offsets of 4, 8, 12, 16, 20, 24, 28, 32 and 36 $\mu$m on each sub-aperture, making none of the fringe peaks at the zero OPD. The final CDR is 7.5 instead of 7.

\section{Conclusions}

We discussed the requirements and limitations of a spatio-spectral recombiner, for a large number of sub-apertures.

We found that 7 configurations exist with the most densely packed fringe peaks, allowing for relatively low spectral resolutions to be used.

These revised configuration provide more densely packed fringe peaks than before, allowing for a gain in spectral resolution and therefore in sensitivity of such an instrument concept.

\acknowledgments{
We thank J. Monnier and M. Ireland for their questions on the instrumental concept and fringe peaks overlap.
}


\end{document}